\documentclass[11pt,twoside]{article}
\usepackage{asp2010}
\usepackage{graphicx}

\resetcounters

\begin{document}

\title{Numerical simulations of relativistic magnetic
  reconnection with Galerkin methods }

\author{O.~Zanotti,$^1$ and M.~Dumbser$^2$
\affil{$^1$Max-Planck-Institut f{\"u}r
  Gravitationsphysik, Albert Einstein Institut, Golm,
  Germany, \\
$^2$Laboratory of Applied Mathematics, University of
  Trento, Via Mesiano 77, I-38100 Trento, Italy
}}

\begin{abstract}
We present the results of 
two-dimensional 
magnetohydrodynamical numerical simulations of
relativistic magnetic reconnection,
with particular emphasis on the 
dynamics of Petschek-type configurations with high Lundquist
numbers, $S\sim 10^5-10^8$.
The numerical scheme adopted, allowing for 
unprecedented accuracy for this type of calculations, 
is based on high order finite volume 
and discontinuous Galerkin methods as recently proposed by
\citet{Dumbser2009}.  
The possibility of producing high Lorentz factors is
discussed, by studying the effects produced  on the dynamics 
by different magnetization and resistivity regimes.
We show that Lorentz factors close to $\sim 4$ can be
produced for a  plasma magnetization parameter $\sigma_m=20$.
Moreover, we find that the Sweet-Parker layers are
unstable, generating secondary magnetic islands,
but only for $S>S_c\sim 10^8$, much larger than 
what is reported in the Newtonian regime.
\end{abstract}

\section{Introduction}
\label{introduction}

Relativistic magnetic reconnection is 
a high-energy process 
converting magnetic energy into heat and plasma kinetic energy
over short timescales. It is supposed to play a
fundamental role in the magnetospheres of
pulsars~\citep{Uzdensky2003,Gruzinov2005}; 
at the termination shock of a relativistic striped pulsar
wind~\citep{Petri2007}; in soft gamma-ray
repeaters~\citep{Lyutikov2003,Lyutikov2006}; 
in gamma-ray burst
jets~\citep{Drenkhahn2002,Barkov2010,McKinney2010,Rezzolla:2011}; 
in accretion disc
coronae~\citep{diMatteo1998,Schopper1998,Jaroschek2004}.

In spite of the initial optimistic expectations that relativistic
magnetic reconnection could provide very fast
reconnection rates, evidence has emerged over the years
that the relativistic Petschek reconnection should
not be considered as a mechanism for the direct
conversion of the magnetic energy into the plasma energy
and the reconnection rate would be at most $0.1$ the
speed of light, contrary to what originally suggested by
\citet{Blackman1994}.
Fundamental progresses in the numerical modeling
of relativistic magnetic
reconnection have been recently obtained by
~\cite{Watanabe2006,Zenitani2009,Zenitani2009_guide}, 
However, two major numerical limitations
still prevent realistic astrophysical applications of
numerical schemes specifically devoted to relativistic
magnetic reconnection.
The first limitation is due to the difficulty in reaching
sufficiently high 
magnetization parameters $\sigma_m$, while the
second limitation is due to the difficulty in treating
physical systems with very high Lundquist numbers $S$. 

In this work, by adopting the innovative numerical method
presented in~\citet{Dumbser2009}, we show the results of 
numerical simulations in the very high Lundquist numbers
regime, $S\sim 10^5-10^8$, showing that 
that Sweet-Parker current
sheets are unstable to super-Alfvenically fast formation
of plasmoid chains, but only for $S>S_c\sim 10^8$.

\section{Physical set up and numerical approach}

The initial model that we have considered 
is built on Harris model, as reported by  \cite{Kirk2003},
and it reproduces a current sheet
configuration in the $x-y$ plane.
Gas pressure and density are given by
$p=p_0+\sigma_m\rho_0[p_0\cosh^2(2x)]^{-1}$, 
$\rho=\rho_0+\sigma_m\rho_0[p_0\cosh^2(2x)]^{-1}$, where $p_0$ and
$\rho_0$ are the constant values outside the current
sheet, whose thickness is $\delta=1$.
The magnetic field changes orientation across the
current sheet according to $B_y=B_0\tanh(2x)$, where 
the value of $B_0$ is given in terms of the magnetization parameter
$\sigma_m=B_0^2/(2\rho_0\Gamma_0^2)$.
All over the grid there is a small background uniform resistivity
$\eta_b$,
except for a circle of radius
$r_{\eta}=0.8$, defining a region of anomalous
resistivity of amplitude $\eta_{i0}=1.0$.
The resistivity can be written as
\begin{eqnarray}
\eta=\left\{
\begin{array}{lr}
{\eta_b +\eta_{i0}\left[2(r/r_{\eta})^3-3(r/r_{\eta})^2+1\right]} & {\rm for} \hspace{12pt} r\leq r_{\eta}, \\
\eta_b & {\rm for} \hspace{12pt} r>r_{\eta}, \\
\end{array}
\right.
\end{eqnarray}
where  $r=\sqrt{x^2+y^2}$.
The velocity field is
initially zero,
while the electric field is given by
$E_z=\eta(\partial B_y/\partial x)$.  
We have considered the case with $p_0=1$, $\rho_0=1$.
The Lundquist
number for every model is $S=v_AL/\eta_b$,
where $L$ is the length of the initial current
sheet, while $v_A^2=B^2/(h\rho+B^2)$ is the relativistic
Alfven velocity.

A well known and challenging feature of 
the relativistic resistive magnetohydrodynamics equations
is that the source terms in the three equations 
for the evolution of the electric field
become stiff  in the limit of high conductivity.
To cope with this difficulty, we have applied the strategy described
by \citet{Dumbser2009}, who used the so called high order 
${P_NP_M}$ methods, which combine high order 
finite volume methods and discontinuous Galerkin finite element
schemes in a more general framework~\citep{DET2008,DBTM2008}.
The numerical grid consists of  an
unstructured mesh composed of triangles 
which are
clustered along the current sheet.
The grid extension is given by
$[-50,50]\times[-150,150]$.
We have used
periodic boundary conditions at $y_{\rm min}$ and $y_{\rm max}$, while
zeroth order extrapolation is applied at $x_{\rm min}$
and $x_{\rm max}$.

\section{Results}

\begin{figure}
\centering
\hspace{0.2cm}
{\includegraphics[angle=0,width=10.0cm,height=4.0cm]{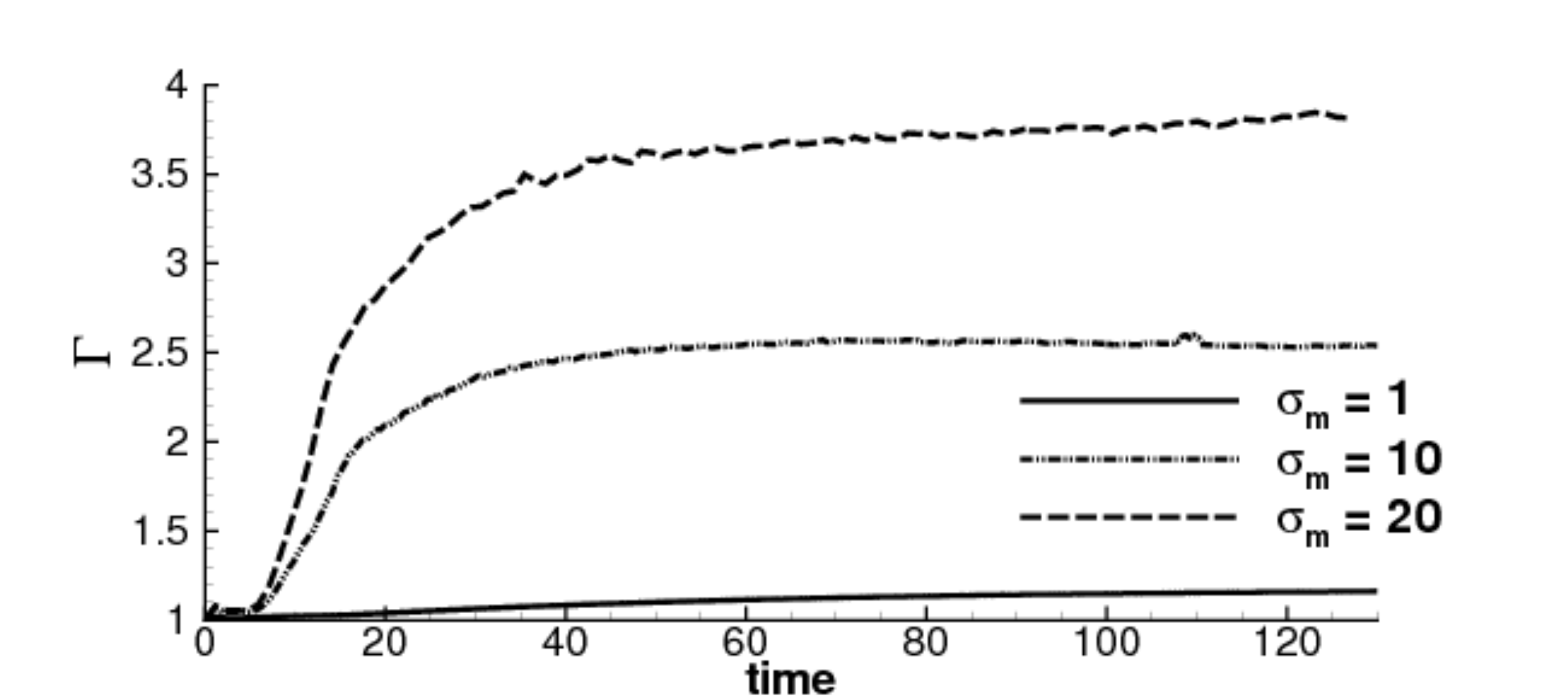}}
\caption{
Time evolution of the 
  Lorentz factor for
  models 
  with increasing magnetizations $\sigma_m$ and $S\sim 10^5$.}
\label{Lorentz_factor}
\end{figure}

The dissipated magnetic energy, triggered by the
anomalous resistivity, produces an increase of
both the thermal and kinetic energy. The
latter results in the acceleration of the plasmoid along the direction of the
magnetic field,
which is more efficient for
higher magnetizations. 
However, as the region around the
anomalous resistivity becomes more and more rarefied, the
conversion of magnetic energy into thermal energy becomes
more efficient than the conversion into kinetic energy
and the Lorentz factor reaches a saturation.
This effect is shown in Fig.~\ref{Lorentz_factor}

\begin{figure}
\centering
\hspace{0.2cm}
{\includegraphics[angle=0,width=4.0cm,height=10.0cm]{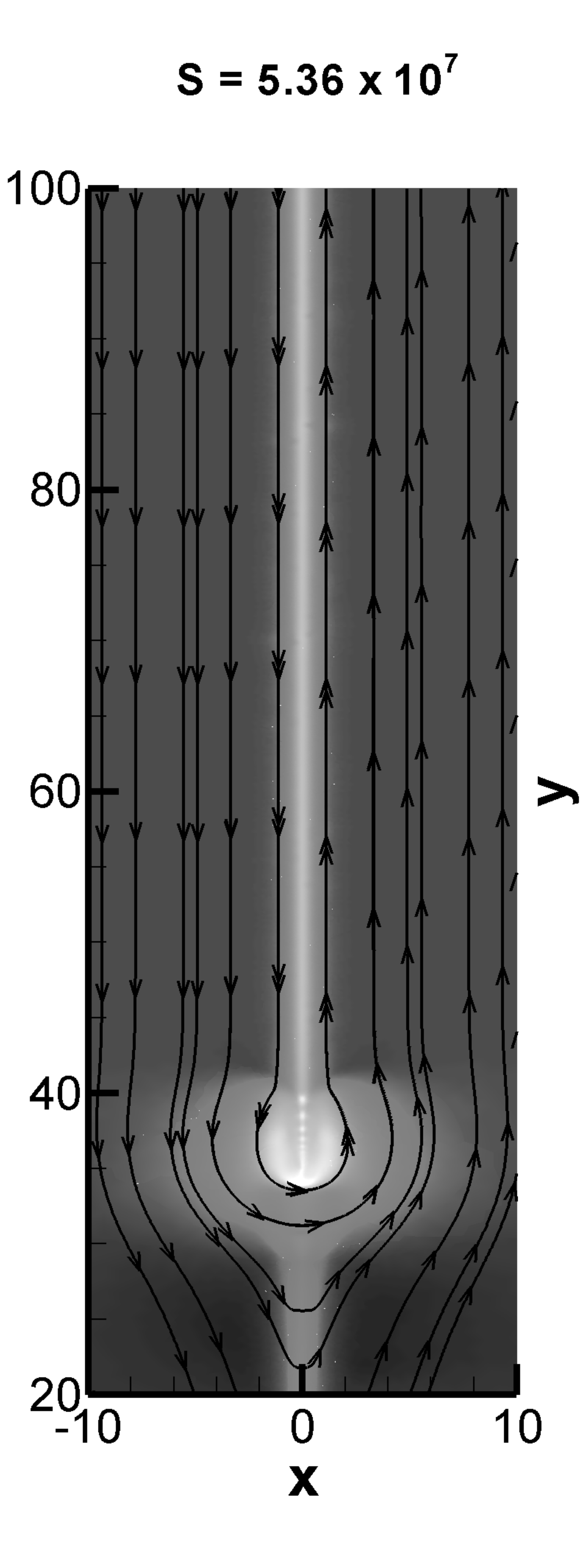}}
{\includegraphics[angle=0,width=4.0cm,height=10.0cm]{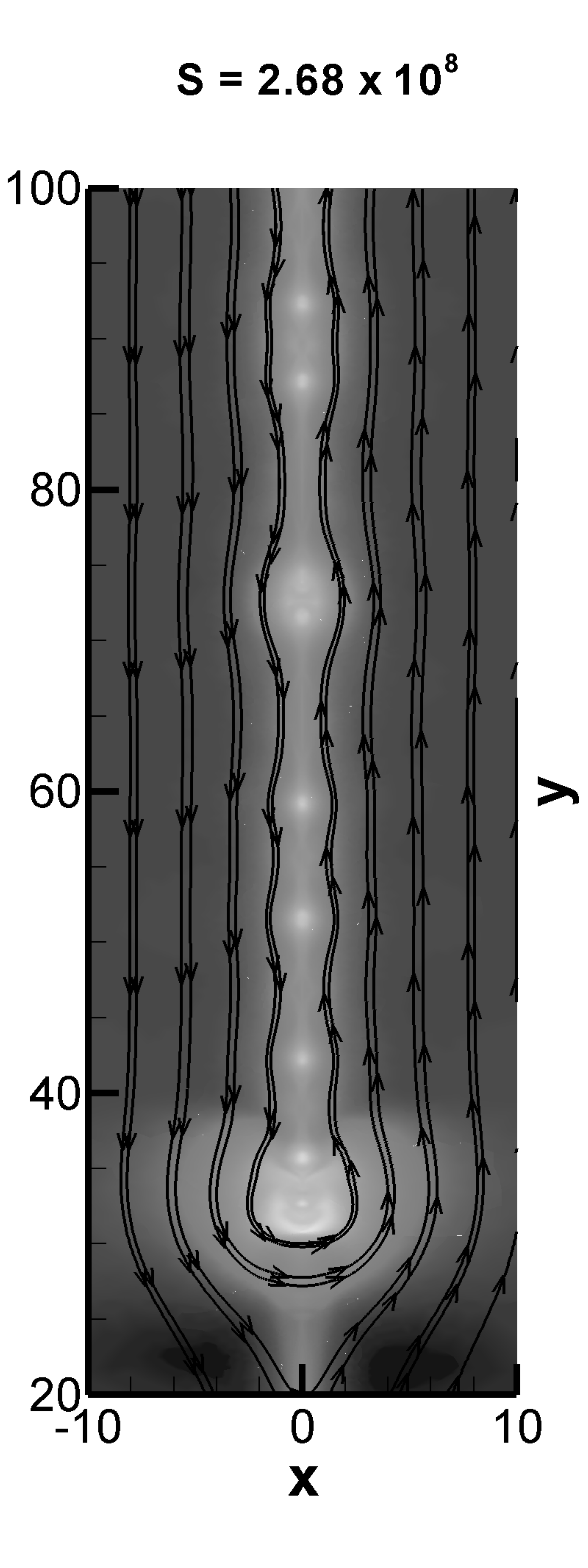}}
\caption{Generation of plasmoid chains in very high
  Lundquist numbers configurations. Left and right
  panels report, respectively,
  the color map of the rest mass density  at time
  $t=60$ for two models having $S=5.36\times 10^7$
  and $S=2.68\times 10^8$.}
\label{plsmoid_chain}
\end{figure}

Fig.~\ref{plsmoid_chain}, on the other hand, shows the
development of an tearing-like instability for very high Lundquist
numbers configurations.
No sign of instability is visible in simulations
with Lundquist numbers as large as $S\sim10^7$ (left panel), while
a chain of magnetic islands is produced for $S>S_c\sim10^8$ (right panel). 
Our results, 
combined with those by \citet{Samtaney2009}, 
indicate that, in
the transition to the relativistic regime, the critical
Lundquist number increases from $S_c\gtrsim 10^4$ to
$S_c\gtrsim 10^8$. Such a conclusion may have deep
implications for high Lundquist number reconnection in
realistic astrophysical conditions, and it requires
further investigations.

\section{Conclusions}

By adopting high order
discontinuous Galerkin methods as proposed by~\citet{Dumbser2009},
we have
found that Lorentz factors up to $\sim4$
can be obtained for plasma parameters $\sigma_m$ up to
$20$ in systems undergoing relativistic magnetic
reconnection 
with Lundquist number $S\sim10^5$.
When $S$ is larger than a
critical value $S_c\sim 10^8$, 
the Sweet-Parker layer becomes
unstable, generating a chain of secondary magnetic
islands~\citep{Zanotti2011}.

\section*{Acknowledgments}
I am grateful to Luciano Rezzolla for useful discussions.
Numerical simulations where performed on the 
National Supercomputer HLRB-II 
installed at Leibniz-Rechenzentrum.


\end{document}